\documentclass[a4paper, final]{article}

\usepackage{INTERSPEECH2016}

\usepackage{amssymb,amsmath,bm}
\usepackage{textcomp}

\def\vec#1{\ensuremath{\bm{{#1}}}}

\ninept

\usepackage[utf8]{inputenc}

\title{A real-time framework for visual feedback of articulatory data\newline
  using statistical shape models}

\makeatletter
\def\name#1{\gdef\@name{#1\\}}
\makeatother
\name{%
  \emph{Kristy James}\textsuperscript{1--2},
  \emph{Alexander Hewer}\textsuperscript{1--3},
  \emph{Ingmar Steiner}\textsuperscript{1--2},
  \emph{Stefanie Wuhrer}\textsuperscript{4}
}

\address{
  \textsuperscript{1}Computational Linguistics \& Phonetics, Saarland University, Germany\\
  \textsuperscript{2}DFKI Language Technology Lab, Saarbrücken, Germany\\
  \textsuperscript{3}Saarbrücken Graduate School of Computer Science, Germany\\
  \textsuperscript{4}INRIA Rhône-Alpes, Grenoble, France\\
  {\small \tt \{kristyj|hewer|steiner\}@coli.uni-saarland.de, stefanie.wuhrer@inria.fr}
}

\usepackage{microtype}

\usepackage{enumitem}

\usepackage{graphicx}

\usepackage[hidelinks]{hyperref}

\usepackage{doi}
\usepackage{IEEEtrantools}

\usepackage[acronym, shortcuts]{glossaries}

\newacronym{ema}{EMA}{electromagnetic articulography}
\newacronym{mri}{MRI}{magnetic resonance imaging}
\newacronym{pca}{PCA}{principal component analysis}

\newcommand*{\keywords}{EMA, articulatory feedback, 3D tongue model}

\newcommand{\eg}{\emph{e.g.}}
\newcommand{\etal}{\emph{et al.}}
\newcommand{\cf}{\emph{cf.}}

\begin{document}
\bstctlcite{IEEEexample:BSTcontrol}

\newsavebox\frameworkfigure
\begin{lrbox}{\frameworkfigure}
  \includegraphics{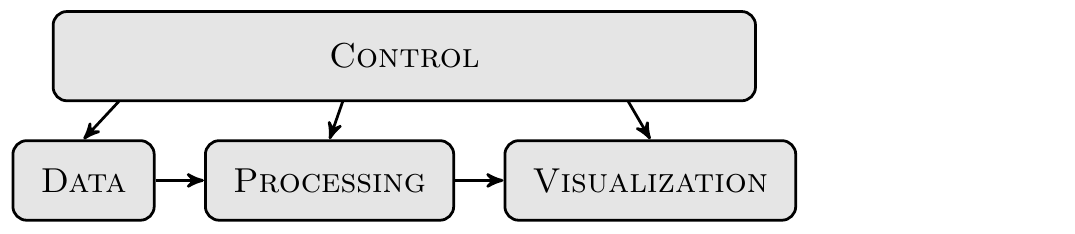}
\end{lrbox}

\maketitle

\begin{abstract}
  We present a novel open-source framework for visualizing \ac{ema} data in real-time,
  with a modular framework
  and anatomically accurate tongue and palate models
  derived by multilinear subspace learning.
\end{abstract}
\glsresetall

\section{Introduction}

Investigating and visualizing the motions of the major articulators is of great interest in speech science.
Such visualizations can be used, for example in speech therapy, to provide feedback for articulation in the form of a virtual talking head.
Ideally, such feedback should occur as fast as possible, which requires a modality that can provide data for visualization with minimal latency.
One such modality is \ac{ema}, which can track the motion of selected points on relevant articulators, such as the tongue tip, in real-time.
However, as only a small number of fleshpoints are tracked, visualizing the acquired data in a meaningful way that reflects the speaker's unique anatomy is a challenging task.

\subsection{Related Work}

Visualizing the vocal tract during speech and providing articulatory feedback is an active field of research:



Badin \etal's Audiovisual Talking Head (ATH) models speech articulators based on \ac{mri} data and video images from one speaker \cite{badin2008audiovisual}. More recent work has focused on animating the ATH using ultrasound data, though to our knowledge this was an offline method \cite{fabre2014automatic}.

Katz \etal's Opti-Speech \cite{Katz2014} uses \ac{ema} to provide real-time articulatory feedback,
using a generic avatar to show the motions of the articulators.
It uses technology that may not be available free of charge; a commercially available system is under development \cite{Vulintus}.

\subsection{Our contribution}

In this paper, we present a cross-platform framework for visualizing \ac{ema} data either played back from pre-recorded data, or streamed live from an articulograph (NDI or Carstens) in real-time, providing articulatory feedback.
The framework is based on open source technology and can therefore be used free of charge.
Furthermore, it is designed in a modular way, which makes it flexible and easy to extend.
Finally, we use statistical models to generate speaker-specific palate and tongue shapes:
for the palate, we utilize a \ac{pca} model, \cf\ \cite{Hewer2015};
for the tongue shape, we integrate a multilinear model.

This paper is organized as follows:
We first outline the statistical models used and how they can be used to visualize the data.
Then we turn to our framework and describe its structure and functionality.
To conclude, we outline future work.

\section{Statistical models}

The statistical models used in this system make use of the following shape representation:
A polygonal mesh \( M := (V, F) \) consists of a vertex set \(V := \{\vec{v}_i\} \) with \(\vec{v}_i \in \mathbb{R}^3\) and a face set \(F\).
A face \(f \in F\) is a set of vertices that form a polygonal surface patch if linked together by edges.

\subsection{Palate model}

In simplified terms, the \ac{pca} palate model \(M_P(\vec{x})\) is a palate shape mesh \(M_P\) that depends on weights \(\vec{x} \in \mathbb{R}^n\).
This weight vector determines the anatomical features of the generated shape.
This model was derived from the \ac{mri} scans of the datasets of Adam Baker \cite{Baker2011} and the Ultrax project \cite{Ultrax2014} by using an approach similar to \cite{Hewer2015}.

We can use this model to reconstruct the palate shape from a palate trace
by finding the best weight \(\vec{x}\) such that \(M_P(\vec{x})\) is close to the data.

\subsection{Tongue model}
\label{subsection:tonguemodel}

The multilinear tongue model \(M_T(\vec{x}, \vec{y})\) is a mesh \(M_T\) that depends on two weights:
\(\vec{x} \in \mathbb{R}^n\) and \(\vec{y} \in \mathbb{R}^m\).
The weight \(\vec{x}\) influences the anatomical features of the tongue, \(\vec{y}\) affects the speech related shape.
The tongue model was obtained from the same datasets as the palate model.

Given an \ac{ema} data sequence, we can utilize this model to generate a dynamic tongue shape as follows:
First, we manually set which vertex of the tongue mesh corresponds to which \ac{ema} coil in the data.
For each frame, we then find the best weights \(\vec{x}\) and \(\vec{y}\) such that the corresponding vertices are near the current coil positions.
Here, we require the weights to be similar to the ones of the previous time step, which results in smooth transitions.
After a few processed samples, \(\vec{x}\) can be set to the average of the previously obtained values in order to fix the anatomical features.
This procedure prevents the optimization process from continuously adapting the tongue anatomy to the received data.

\section{Framework}

\subsection{Overview}

\begin{figure}
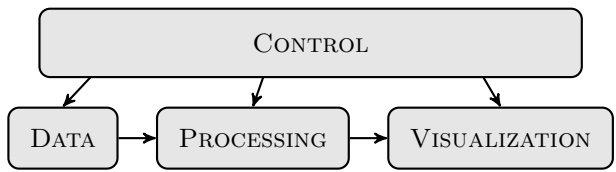

  \centering
  \usebox\frameworkfigure
  \caption{The four units of our framework}
  \label{fig:framework}
\end{figure}

As depicted in \autoref{fig:framework}, our framework can be split into four distinct units, which all can be installed and run on Windows, Mac OSX, and Linux:
\begin{description}[leftmargin=0pt, itemindent=0pt]

  \item {\bfseries Data}
    This unit contains interchangeable modules that are responsible for transmitting \ac{ema} data,
    and associated audio information.
    They may be replaced by an articulograph (currently supporting NDI Wave with potential for others), though
    it is also possible to use pre-recorded data in several formats.

  \item {\bfseries Processing}
    The processing unit serves as a mediator between the data and visualization unit:
    It processes the data before sending it to the visualization unit, for example
    performing head correction and smoothing of the received \ac{ema},
    applying delays or transformations to the data,
    or recording the audio or \ac{ema} signals as they are processed.
    It is here that we find  the optimal weights for the statistical models that approximate the \ac{ema} data.

  \item {\bfseries Visualization}
    This unit renders an intuitive representation of the processed data:
    It visualizes the shape and position of the tongue, lips, lower jaw, and the palate.
    In order to bring the different articulators into context, a generic head shape is also shown.
    Additionally, the unit can augment the generated visualization by also playing the synchronized audio.

  \item {\bfseries Control}
    The final unit includes graphical user interfaces that allow the user to control the different units and to configure the framework.
    For example, the user can decide which \ac{ema} data is used or which processing steps are performed.
    Here, it is also possible to launch specific tasks, such as recording the palate trace or the bite plane of the subject.

\end{description}

\subsection{Implementation details}

Our framework is mostly implemented in Python, which makes it easy to read and extend.
The modules for fitting the statistical models, however, are implemented in C++ for performance reasons, where we use a quasi-Newton solver \cite{Liu1989} to find the optimal weights.
For the visualization, we use the game engine of the open source software Blender \cite{blender}, which supports Python scripting.
Blender has built-in methods for performing inverse kinematics that we use to animate the jaw and lips.
To visualize the tongue and palate, a Python class reconstructs the corresponding mesh from provided weights.
The individual modules of our framework communicate via network protocols, which makes it possible to run individual modules on distributed hardware.

\subsection{Example workflow}

\begin{figure}
  \includegraphics[width=\linewidth]{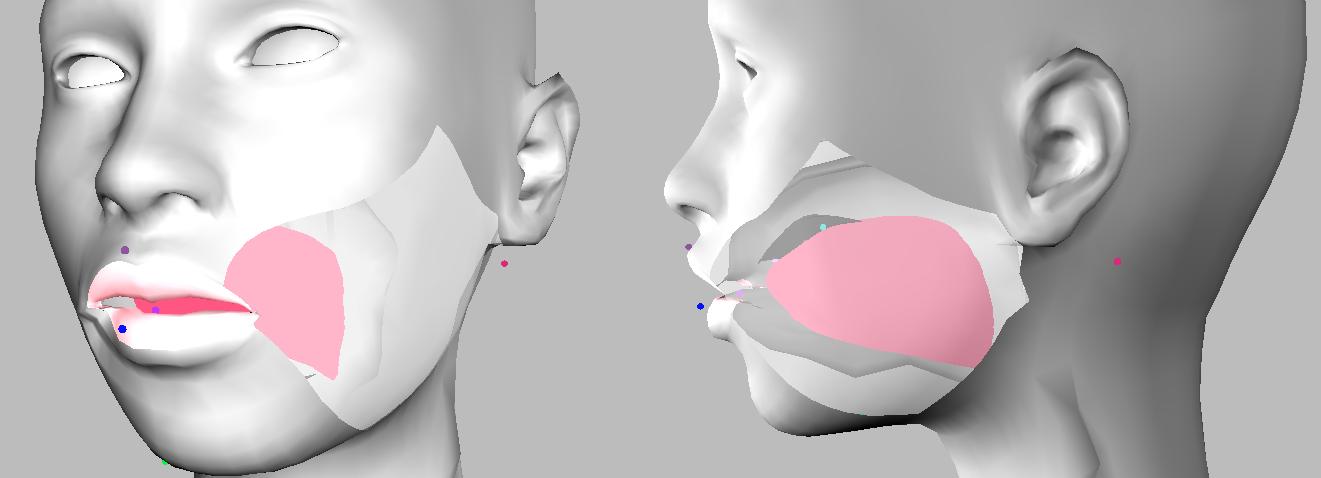}
  \caption{Example visualization}
  \label{fig:example}
\end{figure}

A typical workflow is as follows:
First, the user specifies the roles of the different \ac{ema} coils, \eg, which coils are used for head correction.
This step also includes setting up the correspondences discussed in \autoref{subsection:tonguemodel}.
Next, the bite plane of the subject is recorded, which is used to create the canonical coordinate system the data is represented in.
As the origin of this coordinate system we use a midsagittal point near the upper incisors.
This point can be provided by an \ac{ema} coil or recorded in an additional step.
Next, a palate trace can optionally be recorded that is used to estimate the palate shape by using the statistical palate model.
This concludes the setup, allowing the framework to be used to visualize an \ac{ema} data sequence.
An example visualization created from data of \cite{Steiner2014} is shown in \autoref{fig:example}.

The source code for our framework is made available under a GPL license, and can be found at \url{https://github.com/m2ci-msp/ematoblender}.

\section{Conclusion}

In this paper, we have described a modular, cross-platform \ac{ema} visualization framework that is suitable for articulatory feedback, using open-source software and statistical shape models of the tongue and palate.
Future work includes the integration of teeth into the avatar, as well as conducting a usability study of the framework.

\eightpt
\bibliographystyle{IEEEtran}
\bibliography{references}

\end{document}